\documentstyle[11pt,moriond]{article}

\def\la{\lower.5ex\hbox{$\; \buildrel < \over \sim \;$}}
\def\ga{\lower.5ex\hbox{$\; \buildrel > \over \sim \;$}}

\def\PRL{\em Phys. Rev. Lett.}

\def\ApJ{{\em Astrophys. J.}}

\def\MN{{\em Monthly Not. Roy. Astron. Soc.}}

\def\AA{{\em Astron. \& Astroph.}}

\def\etal{{\it et al.~}}
\def\eg{{\it e.g.,~}}

\def\be{\begin{equation}}
\def\ee{\end{equation}}
\def\bea{\begin{eqnarray}}
\def\eea{\end{eqnarray}}

\begin{document}
\vspace*{4cm}
\title{COSMIC STRUCTURE OF MAGNETIC FIELDS}

\author{Peter L. Biermann$^{1,5}$, Hyesung Kang$^{2,6}$,
J{\"o}rg P. Rachen$^{3,7}$ and Dongsu Ryu$^{2,4,8}$}

\address{$^1$Max-Planck Institute for Radioastronomy, D-53010 Bonn, Germany\\
$^2$Department of Astronomy, University of Washington,
    Box 351580, Seattle, WA 98195-1580, USA\\
$^3$Astronomy Dept, Penn State University, University Park, PA 16802, USA\\
$^4$Department of Astronomy \& Space Science, Chungnam National University,
    Daejeon 305-764, Korea\\
$^5$e-mail: plbiermann@mpifr-bonn.mpg.de\\
$^6$e-mail: kang@hermes.astro.washington.edu\\
$^7$e-mail: jorg@astro.psu.edu\\
$^8$e-mail: ryu@hermes.astro.washington.edu}

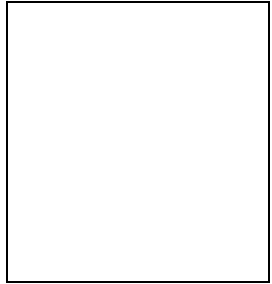
\maketitle
\abstracts{The simulations of the formation of cosmological structure
allows to determine the spatial inhomogeneity of cosmic magnetic fields.  Such
simulations, however, do not give an absolute number for the strength of the
magnetic field due to insufficient spatial resolution.
Combining these simulations with observations of the Rotation
Measure to distant radio sources allows then to deduce upper limits for the
strength of the magnetic field.  These upper limits are of order $0.2 \,
{\rm to} \, 2 \; \mu {\rm gauss}$ along the filaments and sheets of the galaxy
distribution.  In one case, the sheet outside the Coma cluster, there is a
definitive estimate of the strength of the magnetic field consistent with this
range.  Such estimates are almost three orders of magnitude higher than
hitherto assumed usually.  High energy cosmic ray particles can be either
focussed or strongly scattered in such magnetic filaments and sheets, depending
on the initial transverse momentum.  The cosmological background in radio and
X-ray wavelengths will have contributions from these intergalactic filaments
and sheets, should the magnetic fields really be as high as $0.2 \,
{\rm to} \, 2 \; \mu {\rm gauss}$}

\section{Introduction}

The observation of the rotation of the plane of polarization as a function of
wavelength in extragalactic radio sources has long been known to give unique
information on the strength of cosmological magnetic fields (Kronberg 1994).
This technique gives information out to redshift of about 2.5, and can in
principle do this on the line of sight to a polarized radio source at any
redshift.

Information about the strength of cosmological magnetic fields is of key
importance for our physical understanding of the early universe.  In the
standard model for the early phases of the universe there may be a possibility
to create magnetic fields, but the data do not require this.  It is well 
known that there is a possibility to create weak magnetic fields form zero 
by the Biermann battery (Biermann 1950) in a turbulent system with net 
helicity, such as a rotating star or a shock wave can give.  Building on 
this mechanism several possible origins of the observed magnetic fields 
can be found; we will discuss two approaches here.

The data situation with respect to cosmic magnetic fields is as follows
(Kronberg 1994):

We find neutron stars with magnetic fields of up to about $10^{12}$ gauss,
possibly even more, white dwarfs with fields up to about $10^8$ gauss, and
normal stars such as the Sun with localized fields up to several $10^3$
gauss.  The origin of the solar magnetic field is believed to be the lower
layers or the lower boundary zone of the hydrogen convection zone on the Sun.
This convection zone extends all the way to the surface of the Sun as in all
low mass stars, and so the magnetic field can be transported to the
photosphere.  Interestingly, also massive stars such as OB or WR stars with
radiative envelopes show clear evidence of magnetic fields through
non-thermal synchrotron emission; the magnetic field may have been
transported through circulation currents induced by rotation from the
convective core region to the surface.

In the interstellar medium of our Galaxy as well as other galaxies (Beck \etal
1996) the magnetic field is of order $10^{-5}$ gauss, with higher values
possible in starburst galaxies such as M82.  Even at high redshift already
galaxies have their normal magnetic field.  This magnetic field appears to be
usually in equipartition with the thermal matter in the galaxy, even for
short-lived starburst phases of galaxies such as M82.  This requires a
mechanism to create and/or strengthen the magnetic field which is fast, active
over just a few rotation periods of a galaxy.

In clusters of galaxies there is evidence for magnetic fields as well; 
however, due to the uncertainty of the reversal scale of the field, {\it i.e.}
the scale over which the direction changes sign, the strength of the
intra-cluster fields is not clear; it may be a few $10^{-6}$ gauss or 
even nearer to $3\times 10^{-5}$ gauss, again near equipartition (En{\ss}lin
\etal 1997).

Outside clusters of galaxies there is one estimate of a magnetic field, again
from Kronberg's group (Kim \etal 1989), of order $3 \times 10^{-7}$ gauss, in
a filament outside the Coma cluster.  We will come back to this measurement
below.

Finally, across the universe, one uses the variance of the measured Rotation
Measures along the line of sight to many radio quasars at high redshift.
The observational Rotation Measure (RM) data of radio quasars limit RM to
$$
RM \, \sim \, 5~{\rm \, rad~m^{-2}}
$$
out to a redshift of $z=2.5$ (Kronberg \& Simard-Normandin 1976;
Kronberg 1994 and references therein).

In order to use this measurement, one needs a model for the geometry of the
magnetic field.  Assuming that the magnetic field is basically homogeneous in
a comoving volume, but reverses its direction every length $L_{\rm rev,Mpc}$
(the reversal scale in Mpc), gives
$$
B_{IGM} \la 10^{-9} L_{\rm rev,Mpc}^{-1/2} \; {\rm gauss}.
$$
If we were to use instead of 1 Mpc for the scaling the bubble scale of the
galaxy distribution, then obviously, this upper limit to any magnetic field
would be much lower, more like $2\times 10^{-10}$ gauss.

In this paper we derive a model for the cosmological geometry of the magnetic
field, and then use the data to derive new upper limits for the strength of
the magnetic field in the cosmological structures.  What we will show is that
this exercise leads to an estimate which is about three orders of magnitude
higher.

Some of the points described here have been developed in more detail in
Biermann, Kang, \& Ryu (1997) and Ryu, Kang \& Biermann (1997).

We use in the following $h_{0.5}$ as a measure of $H_o$ in units of $50~{\rm
km/s/Mpc}$. In section 4, ultra high energy cosmic ray energies are given in
$\rm EeV = 10^{18}\,eV$.

\section{The magnetic field in the cosmos}

Our normal understanding of the structure formation in the universe starts with
gravitational instabilities, which lead to the formation of Zeldovich pancakes,
which in turn intersect with each other and so produce the large scale
distribution of galaxies, which we observe.
This distribution can be described as a network of
irregular soap-bubbles, with filaments arising from the intersection of
bubble-walls, and clusters of galaxies at the triple point of
intersection.  These models then also lead to a steady
accretion towards the structures and to shocks around them (\eg Ryu \etal
1993;  Kang \etal 1994; Ryu \& Kang 1997b).

If there was a magnetic field already before the first galaxies formed, then
these streaming motions and shocks would strengthen the magnetic field
(\eg Kulsrud \& Anderson 1992; Kulsrud \etal 1997).  If the magnetic field
formed with the formation of structure, as suggested by Kulsrud \etal (1997),
then also todays distribution would strongly correlate with the galaxy
distribution.  Finally, if the magnetic field formed along with the first
galaxies, but form stellar activity, again we would expect the same
cosmological inhomogeneity.  The consequences of this is what we explore here.

In the following we explore the these two models in some detail.

\subsection{Field Generation via Large Scale Structure Formation}

In the structure formation of the universe shocks develop along the sheets,
filaments and clusters, with strong shearing motions (\eg Ryu \etal 1993;
Kang \etal 1994; Ryu \& Kang 1997b).  The shock velocities around clusters
reach about
$$
v_{acc} \, \approx \, 10^8 \, \rm cm/s \,
(M_{cl}/R_{cl})/(4 \times 10^{14}{\rm M}_{\odot}/{\rm Mpc})]^{1/2}
$$
for a range of possible densities of the universe of
$0.1 \, \le \, \Omega_o \, \le \, 1$ (Ryu \& Kang 1997a).

If there was no primordial field, then the Biermann battery mechanism can
produce a weak magnetic field in these shocks.  This has been followed
through numerical simulations in cosmological contexts (Kulsrud \etal 1997;
Ryu, Kang \& Biermann 1997).  The turbulent cascade which develops, a
Kolmogorov cascade, then pushes the magnetic field energy up in scale to
produce a large scale magnetic field in clusters of galaxies (Kulsrud \etal
1997).  The inverse cascade pushes the magnetic field then up to levels of
order $10^{-5}$ gauss, consistent with observations (En{\ss}lin \etal 1997).

This kind of simulations does not have sufficient spatial resolution to follow
all the small scale motions in the filaments and sheets, but we conjecture
that the field would also be increased.  However, the simulations show that
the magnetic field, embedded in the fully ionized intergalactic medium, is
indeed strongly correlated with the galaxy distribution, and so is expected
to be much stronger along filaments and sheets than in the voids.

\subsection{Field Generation via Stellar Dynamos and Subsequent Expulsion}

The Sun operates a dynamo, changing its magnetic field in polarity every 11
years.  This is interpreted as the consequence of a dynamo, which is now
believed to operate at the lower boundary zone of the hydrogen convection
layer.  Other stars clearly can also produce magnetic fields, and so the
observation of ubiquitous magnetic fields on stars can be understood.  The
massive stars also have evidence for magnetic fields
as seen through non-thermal
radio emission (Abbot \etal 1986; Bieging \etal 1989).
Massive stars explode as supernovae, then spilling their magnetic field into
the interstellar medium.  The most massive stars already pollute the
interstellar medium with their magnetic field through their powerful winds.
These winds have been argued to contain magnetic fields up to a an order of
magnitude of $3\times 10^3$ gauss on the surface of the star (Maheswaran \&
Cassinelli 1992).  Then these winds may actually have some additional momentum
due to their magnetic fields (Seemann \& Biermann 1997).

If the moderate estimates for the strength for these magnetic fields in stellar
winds (Biermann \& Cassinelli 1993) can be confirmed, then the winds terminate
in the interstellar medium with a remaining field of order $10^{-5}$ gauss
(after allowing for a factor of 4 from the shock transition).  This means that
the magnetic field injected into the interstellar medium is already near
equipartition from the injection region, which changes the requirements for a
successful dynamo theory drastically.  The dynamo process no longer has to
strengthen the field, but it has to order it starting from near equipartition.
Such ordering has been found in simulations (A. Shukurov, personal
communication through
R. Beck to P.L. Biermann), and so this would enable the dynamo process to
operate on a few rotation periods, rather than very many rotation periods,
which are not available for a galaxy.

This mechanism would also allow to explain why starburst galaxies have such a
strong magnetic field, but again in equipartition with the thermal energy
density of the matter (Kronberg \etal 1985), despite the fact that a typical
starburst cannot be very old, at most $10^8$ years.

A similar argument can be made for the battery mechanism and the dynamo
operating in compact accretion disk, be it around protostars, white dwarfs,
neutron stars, or black holes.  Especially the black holes now believed to lurk
in every early Hubble type galaxy are often observed to eject jets that power
gigantic radio lobes.  These jets and lobes are filled with a magnetic field,
as testified once again by the non-thermal synchrotron radio emission.

\section{Application of the models}

In the simulation we used to estimate the magnetic field strength in
filaments and sheets, a standard cold dark matter (CDM) model was adopted
with $\Omega_b \, = \, 0.06$, $h \, = \, 0.5$, and a bias parameter of
$\sigma_8 \, = \, 1.05$ (see Kulsrud \etal 1997; Ryu, Kang \& Biermann 1997).
With a large number of randomly selected straight lines through a simulation
we then calculated the Rotation Measure through such a simulated universe,
and its variance to a source at redshift $z \, = \, 2.5$.  Care was taken to
avoid clusters of galaxies, because we do not propose to discuss their
magnetic field here; however, it follows already that the fairly high
magnetic fields advocated by En{\ss}lin \etal (1997) are consistent with the
simulations.  Comparing the variance of the Rotation Measure in the
simulations with the observed upper limit we then obtain an upper limit of
the magnetic field strength in filaments and sheets.

Depending on the various cosmology models, we have a range of possible upper
limits $B_u$.  At the observed upper limit of the Rotation Measure we so find
$$
0.2 \, \mu{\rm gauss}~h_{0.5}^{-2} \; < \; B_u \; <
\; 2 \, \mu{\rm gauss}~h_{0.5}^{-2}~.
$$
It is then interesting to come back to the one case where a definite estimate
for an extra-cluster magnetic field has been made, by Kim \etal (1989).  They
find magnetic fields in the plane of the Coma/A1367 supercluster of 0.3 to
0.6 $\mu$gauss, which is already in the range, but still consistent with our
upper limits.

\section{Conclusion and Discussion}

In conclusion, our result indicates that the present limit set by RM
observations allows for the existence of magnetic fields of up to $\sim
1~\mu{\rm gauss}$ in filaments and sheets, if the magnetic field in the voids
is much smaller. Such a geometry of the universal magnetic field is expected
by simulations of structure formation. We shall discuss some consequences of
this result:

\subsection{Equipartition fields and cosmological background radiation}

We may compare the above upper limit with the strength of the magnetic field
whose energy is in equipartition with the thermal energy of the gas in
filaments and sheets. The equipartition magnetic field strength can be
written as,
$$
B \; = \; 0.17 \, h_{0.5} \, \sqrt{{T\over3\times10^6{\, \rm K}}
\,{\rho_b\over{0.3 \, \rho_c}}} \, \mu{\, \rm gauss}
$$
(Ryu, Kang, \& Biermann 1997). The fiducial values $T = 3\times10^6 \,{\rm K}$
and $\rho_b = 0.3 \, \rho_c$ may be appropriate for filaments (see, Kang \etal
1994), but may be somewhat too large for sheets. So, the predicted
equipartition field strength in filaments is $\sim 0.3\,\mu{\rm gauss}$, and
maybe $\la 0.1\,\mu{\rm gauss}$ in sheets, which is close to, but several
times smaller than the overall limit set by RM observations.

Filaments and sheets can contribute radiation at X-ray wavelengths to the
cosmological background. Using fiducial values and a typical size of $20
h_{0.5}^{-1}$Mpc, the luminosity of filaments and sheets in the soft X-ray
band ($0.5{-}2\,$keV) is expected to be $\la 3\times 10^{41}\,{\rm
erg\,s^{-1}}$.  Actually, Miyaji \etal (1996) and Soltan \etal (1996) reported
the detection of extended X-ray emission from structures of a comparable size
and a luminosity of $\approx 2.5 \times 10^{43}\,{\rm erg\,s^{-1}}$, which is
correlated to Abell clusters, suggesting that the temperature and gas density
outside clusters could even be considerably larger than the fiducial values
assumed above. In any case, if this observation can be further confirmed, and
equipartition of magnetic field and gas density is assumed, it would imply
the existence of a $\sim 1\,\mu{\rm gauss}$ magnetic field on large scales
{\em outside} galaxy clusters, and thus be an impressive confirmation of our
results.

Another interesting exploration is the synchrotron radio emission arising
from filaments and sheets of a magnetic field of order $0.1{-}1\;\mu{\rm
gauss}$.  We noted that the original estimate of the magnetic field in the
plane of the Coma/Abell 1367 supercluster was based on a synchrotron radio
continuum measurement (Kim \etal 1989). Since our RM upper limits for the
magnetic field refer to a gross average of filaments and sheets in the
universe, the comparison to just a single observation is likely to be
affected by selection effects. If such estimates could be made for a
statistically complete sample of locations outside big clusters, this would
constitute another definite test of the underlying concept in the simulations
described above, including the strength of the magnetic field more directly
than X-ray observations. Since this project would involve observations at low
radio frequencies, it would be ideal for the new capabilities coming on line
at the GMRT (Giant Meterwave Radio Telescope) in India and at the VLA in the
USA.

Whether filaments and sheets can contribute radiation at other wavelengths,
such as $\gamma$-ray energies, to the cosmological background remains an
unanswered but challenging question at this time.

\subsection{Origin of ultra high energy cosmic rays}

The discoveries of several reliable events of high energy cosmic rays at an
energy above $100\,{\rm EeV}$ raise questions about their origin and path in
the universe, since their interaction with the cosmic microwave background
radiation (MBR) limits the distances to their sources to less than 100 Mpc,
perhaps within our Local Supercluster.

The large-scale accretion shocks in reagions where a seed magnetic field can
be generated are probably the biggest shocks in the universe, with a typical
size in the range of 1-10 Mpc and very strong with a typical accretion
velocity $\ga 1000\,{\rm km}\,{\rm s}^{-1}$.  With up to $\sim 1\,\mu{\rm
gauss}$ magnetic field around them, they could serve as possible sites for
the acceleration of high energy cosmic rays by the first-order Fermi process
(Kang, Ryu \& Jones 1996). With the particle diffusion model in
quasi-perpendicular shocks (Jokipii 1987), the observed cosmic ray spectrum
between $10$ and $100\,{\rm EeV}$ could be explained with reasonable
parameters if about $10^{-4}$ of the infalling kinetic energy can be injected
into the intergalactic space as energetic particles; for the more
conservative assumption of Bohm diffusion, however, no considerable
contribution above $10\,$EeV is expected (Kang, Rachen \& Biermann 1997).

It has been speculated whether the shocks on larger scales, like expected
around filaments and sheets, could make a significant contribution as well
(Norman, Melrose \& Achterberg 1995); a closer look shows, however, that due
to Bethe-Heitler losses at the microwave background this can work out only up
to $\sim 30\,$EeV, and only if quasi-perpendicular shocks are
assumed. However, the simulations suggest that the latter assumption may
apply, and even though filaments and sheets are still not expected to be
efficient accelerators of the highest energy cosmic rays, the special shock
geometry present around them can have a severe impact on their propagation,
as discussed below.

\subsection{Propagation of ultrahigh energy cosmic rays}

The Haverah Park and Akeno data indicate that the arrival directions of the
ultra high energy cosmic ray particles are in some degree correlated with the
direction of the Supergalactic Plane (SGP; Stanev \etal 1995; Hayashida \etal
1996; Uchihori \etal 1997). It was pointed out by Waxman \etal (1997), that
this correlation appears to be better than the correlation of any reasonable
source population to this sheet.

This apparent contradiction is removed by the assumption of a magnetic field
$\la 0.1\,\mu{\rm gauss}$ aligned with the SGP. It is easy to show that
cosmic rays are confined in the sheet if $E<e B_{\rm sh} H_{sh} \sim
700\,{\rm EeV}$ for the fiducial values used above and a sheet thickness of
$5/h_{0.5}\,$Mpc; thus confinement is possible up to the highest observed
cosmic ray energies without exhausting the limits allowed in our model. This
involves essentially two effects which together pronounce the supergalactic
plane in the cosmic ray arrival direction distribution: (a) Due to the
accretion flows infalling toward the supergalactic plane, the high energy
cosmic can be focussed in the direction perpendicular to the SGP, analogously
but in the opposite direction to solar wind modulation.  This ``focussing"
means that particles with a sufficiently small transverse initial momentum
would slowly decrease their transverse momentum even more by interaction
with the incoming accretion flow; this then produces transverse
trapping.  This effect would be most efficient for those particles with a
sufficiently small initial momentum transverse to the sheet (Biermann, 
Kang \& Ryu 1997). However, the details of this effect depend on the 
magnetic field configuration and need further consideration. (b) Since the
particles are captured inside the sheets, the dilution with distance $d$ 
is $1/d$ instead of $1/d^2$, increasing the cosmic ray flux from any source
within the direction of the SGP appreciably compared to the 
three-dimensional dilution expected for other sources. So we may see 
sources in the SGP to much larger distances than expected so far.  On 
the other hand, most particles, {\it i.e.} those with a large initial 
transverse momentum, would be strongly scattered at the  magnetic field,
obliterating all source information from their arrival  direction at Earth,
consistent with the data (Uchihori \etal 1997, Waxman \etal 1997).

Another effect influences the modification of the cosmic ray spectrum at the
highest observed energies, where interactions with the MBR limit the
propagation distance to $\la 100\,$Mpc. Here, the presence of shock waves
around the filaments and sheets leads to a truly universal particle
acceleration, competing with the energy losses. Since the particles never
effectively leave the acceleration region, the advection losses present in
normal shock acceleration scenarios are replaced by the energy losses in the
MBR, leading to a stationary spectrum comparable with a $E^{-3}$ spectrum, if
the acceleration timescale is larger than the loss time scale by about a
factor of 4 (Rachen 1997). Superposed with a flat injection spectrum of point
sources (like radio galaxies of clusters of galaxies), we can expect that the
cosmic ray spectrum can continue with the observed slope up to more than
$300\,$EeV, even if the closest source is more (but not much more) distant
than $\sim 100\,$Mpc. Also here, more detailed calculations are required to
discern the real limits, but it is already clear that most results obtained
for the extragalactic transport of cosmic rays, which were obtained without
regarding the large scale structure, have to be reconsidered.

\section*{Acknowledgments}

We are grateful to Dr. Kronberg for extensive comments on the manuscript.  We
also thank Dr. Beck for comments on the issues raised here.  The work by DR
and HK was supported in part by NASA HPCC/ESS in University of
Washington. The work by DR was supported in part by Seoam Scholarship
Foundation in Chungnam National University. Work of JPR is funded by NASA
grant 5-2857.

\end{document}